\documentclass[aps,prd,preprint,superscriptaddress,amsmath,amssymb,showpacs]{revtex4-1}
\usepackage{dcolumn}
\usepackage{graphicx}
\usepackage{float}
\usepackage{physics}
\usepackage[colorlinks=true,allcolors=blue]{hyperref}

\begin{document}

\title{ The potential of QQQ in the anisotropic background}

\author{Jing Zhou}
\email{zhoujing@hncu.edu.cn}
\affiliation{Department of Physics, Hunan City University, Yiyang, Hunan 413000, China}
\affiliation{All-solid-state Energy Storage Materials and Devices Key Laboratory of Hunan Province, Hunan City University, Yiyang 413000, China}

\author{Kazem Bitaghsir Fadafan}
\email{bitaghsir@shahroodut.ac.ir}
\affiliation{Faculty of Physics, Shahrood University of Technology, P.O.Box 3619995161 Shahrood, Iran}

\author{Xun Chen}
\email{chenxun@usc.edu.cn}
\affiliation{School of Nuclear Science and Technology, University of South China, Hengyang 421001, China}

\begin{abstract}
In this work, we use the AdS/CFT correspondence to study the behavior of a triply heavy baryon within anisotropic backgrounds. Beginning with the total action of the three quarks, we derive the balance equation for the three-quark system and compute the separation distance and potential energy. Our results reveal a consistent decrease in both the separation distance and potential energy for the A configuration and the B configuration as the anisotropy coefficient $a$ increases. This suggests that the presence of an anisotropic background promotes the dissolution of the three-quark system. Additionally, we compare the potential energies of the A and B configurations and observe that the A configuration has a slightly smaller potential energy, suggesting greater stability compared to the B configuration.
\end{abstract}
\maketitle
\section{Introduction}\label{sec:01_intro}
The triply heavy baryons, comprising the three heaviest quarks within the standard model, serves as a valuable way for unraveling the intricate structure and dynamics of baryons. The complications of light-quark interaction are absent in the triply heavy baryons, thus,
they provide an ideal place for our better understanding the
heavy quark dynamics \cite{Liu:2019vtx}. It also can provide information about the quark confinement mechanism
as well as elucidating our knowledge about the nature of the strong force by providing a clean probe of the interplay
between perturbative and nonperturbative QCD \cite{Brambilla:2010cs,Padmanath:2013zfa}.

The three-quark potential is one of the most important inputs of the potential models \cite{Andreev:2015riv,Jiang:2022zbt,Jiang:2023lmj}.However, despite the progress made in studying the triply heavy baryons, there is currently no precise formula available to three-quark potential.
The best known phenomenological models are ansatze which are a kind of the Cornell model for a three-quark system \cite{Andreev:2015riv,Jiang:2022zbt,Jiang:2023lmj}. The short-range part of the potential energy in system can be approximated by the Coulomb potential, which represents the impact of single gluon exchange within the framework of perturbative QCD \cite{Mathieu:2007fpv}. However, in the long-distance part, this potential energy exhibits a linear behavior. Therefore, the potential energy of three quarks can be considered as the sum of the Coulomb potential resulting from interactions between two quarks, and a linear potential arising from the string tension \cite{Takahashi:2002bw,Alexandrou:2001ip,Alexandrou:2002sn,Vijande:2014uma}.

The QQQ potential is a key quantity to understanding the quark confinement mechanism in baryons \cite{Andreev:2015riv}. The difficulty in studing quark confinement is due to the non-perturbative features\cite{Jiang:2022zbt}. The non-perturbative nature of this problem can be effectively addressed by the AdS/CFT correspondence \cite{Maldacena:1997re,Gubser:1998bc,Witten:1998qj}. Subsequently, this correspondence method has been widely utilized for studying heavy quarks. For example, in Ref.\cite{Zhou:2020ssi}, the authors studied the potential energy of heavy quarks in a magnetic field background. Ref.\cite{Ewerz:2016zsx} studied free energy of a heavy quark-antiquark pair in a thermal medium. Recently, charmed pentaquarks and tetraquarks have been studied in \cite{Sonnenschein:2024rzw}. It was found that the presence of angular velocity can facilitate the transition of quark pairs from confinement to deconfinement in a rotating background \cite{Zhou:2021sdy}. In addition, there are some studies on quark pairs under other conditions \cite{Tahery:2022pzn,Wu:2022ufk,Kioumarsipour:2021zyg,MartinContreras:2021bis,Zhang:2020upv,Zhao:2019tjq,Bellantuono:2017msk,Braga:2016oem,BitaghsirFadafan:2015zjc,Iatrakis:2015sua,Hashimoto:2014fha,Finazzo:2014rca,Hou:2007uk}. The above research only focuses on two-quark systems and hasn't involved the QQQ system. Recently, with the proposal of Andreev's series of works, it has become possible for us to study three quark systems from a holographic model \cite{Andreev:2015riv,Andreev:2008tv,Andreev:2020pqy}.

Previous studies of QQQ with isotropic background concentrated on three facets: spectroscopy, production and decay \cite{Wang:2018utj,Brown:2014ena,Can:2015exa,Meinel:2012qz,Yang:2019lsg,Vijande:2015faa,Zhang:2009re,Yin:2019bxe,Martynenko:2007je,Wei:2016jyk,Serafin:2018aih,GomshiNobary:2004mq,Flynn:2011gf,Wang:2018utj}. While less attention has been paid to the anisotropic background. The anisotropy is a consequence of distinct pressure gradients, leading to the rapid expansion of QGP predominantly in the longitudinal direction. Additionally, the isotropic model is unable to replicate the observed energy-dependent behavior of experimental multiplicity, while the anisotropic model effectively captures this dependence \cite{Arefeva:2018hyo,Strickland:2013uga}. Several works have been initiated to study anisotropic backgrounds. For instance, in Ref.\cite{Arefeva:2022avn}, the authors examine the behavior of light quarks within anisotropic backgrounds, while in Ref.\cite{Prakash:2023wbs}, the focus is on investigating heavy quark radiation within an anisotropic hot QCD medium. Additionally, Refs.\cite{Gursoy:2020kjd, Janik:2008tc,Sajadi:2023ckp,Fadafan:2012qu,Shukla:2023pbp,Arefeva:2021jpa,Kasmaei:2018oag,Thakur:2012eb,Prakash:2021lwt,Mateos:2011ix,Giataganas:2012zy,Giataganas:2017koz,Ge:2014aza} contains other notable works addressing anisotropic backgrounds.
Motivated by this, we mainly study the three quark potential energy in anisotropic backgrounds. Specifically, it is necessary to know how the potential energy of three quarks varies with the anisotropic background.

The work is organized as follows: In Sec.~\ref{sec:02_Three-quark potential of A model}, the primary focus is on the potential energy of A configuration. It begins with an examination of the action of three quarks, from which we derive the separation distance and potential energy by establishing the balance equation. Similarly, Sec.~\ref{sec:03_Three-quark potential of B model} extends this discussion to the potential energy in the B configuration. In Sec.~\ref{sec:04}, the primary emphasis is on studying the quark potential energy and separation distance under the same entropy density. The final section serves as a summary of the preceding discussions.
\section{Three-quark potential of A configuration}\label{sec:02_Three-quark potential of A model}
 \begin{figure}
\centering
    \resizebox{0.6\textwidth}{!}{
    \includegraphics{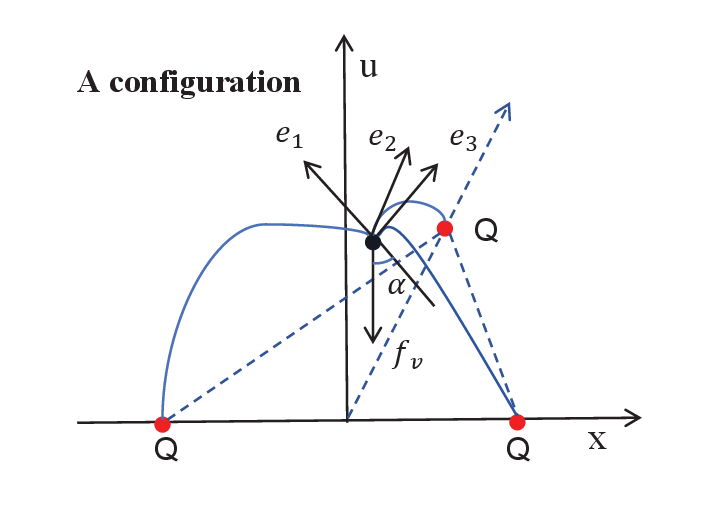}}
    \caption{\label{A} The schematic diagram of A configuration. Red dots are employed to represent heavy quarks, black dot is used to denote verter, and solid blue lines are utilized to illustrate strings. }
\end{figure}
Based on Refs.\cite{Andreev:2015riv,Jiang:2022zbt,Jiang:2023lmj}, our study investigates baryons comprised of three fully heavy quarks(QQQ) in anisotropic background. We have identified two distinct quark configurations for this system: the A configuration and the B configuration. In this section, we concentrate on the A configuration, where two heavy quarks are symmetrically positioned on opposite sides of the x-axis, with the third quark located along the y-axis, as shown in Fig.\ref{A}.

Using the AdS/CFT correspondence, the anisotropic backgrounds have been studied in \cite{Mateos:2011ix, Janik:2008tc}. By deforming the gauge theory with $\theta$ term which depends on one of the boundary field theory coordinates, the anisotropic supergravity solution has been found in \cite{Mateos:2011ix}. Let's start with the string framework of this anisotropic background
\begin{eqnarray}
d s^{2} = \frac{1}{u^{2}}\left(-\mathcal{F} \mathcal{B}\, d t^{2}+d x_{}^{2}+d y_{}^{2}+\mathcal{H} d x_{3}^{2}+\frac{d u^{2}}{\mathcal{F}}\right)+ e^{\frac{1}{2}\phi} d \Omega_{S^{5}}^{2}, \\
\chi = a\, x_{3}, \quad \phi = \phi(u).
\end{eqnarray}
Here $a$ is the anisotropic parameter with units of inverse length, $\phi$ is the dilaton field, $\chi$ is axion which linearly dependents on the coordinates of $x_{3}$. The orientation along the $x_{3}$ axis signifies the anisotropic direction. The anisotropy is attributed to the hydrodynamic dynamics expansion of QGP. Meanwhile, the two directions($x$ and $y$) exhibiting rotational symmetry within the transverse plane are indicative of the transverse direction \cite{Giataganas:2012zy}. Typically, the pressure along the $x_{3}$ direction is less than that in the transverse direction.
The functions $\mathcal{F}(u)$, $\mathcal{B}(u)$, and $\mathcal{H}(u)$ depend on the anisotropy parameter $a$. One finds the isotropic background at $a=0$ so that $\mathcal{F}(u)= \mathcal{B}(u)=\mathcal{H}(u)=1$ and $\phi(u)=0$. Then the temperature can be expressed as
	\begin{equation}
		T = -\left.\frac{\partial_{u} \mathcal{F} \sqrt{\mathcal{B}}}{4 \pi}\right|_{u = u_{h}} = \frac{1}{\pi u_{h}}+a^{2} u_{h} \frac{5 \log 2-2}{48 \pi}+\mathcal{O}\left(a^{4}\right).
	\end{equation}
	The entropy density is given by \cite{Mateos:2011ix}
	\begin{eqnarray}
		s & = & \left(\frac{\pi^{2} N_{c}^{2}}{2}\right) \frac{e^{\frac{-5}{4}} \phi_{h}}{\pi^{3} u_{h}^{3}}.\label{s}
	\end{eqnarray}
To compare results with the isotropic case, we can fix either temperature or entropy in the system. For small $a/T$, the functions $\mathcal{F}(u)$, $\mathcal{B}(u)$, and $\mathcal{H}(u)$ are given by \cite{Mateos:2011ix}
\begin{eqnarray}
\mathcal{F}(u) & = & 1-\frac{u^{4}}{u_{h}^{4}}+a^{2}\mathcal{F}_{2}(u)+\mathcal{O}\left(a^{4}\right), \\
\mathcal{B}(u) & = & 1+a^{2} \mathcal{B}_{2}(u)+\mathcal{O}\left(a^{4}\right), \\
\mathcal{H}(u) & = & e^{-\phi(u)}, \quad \text {with} \quad \phi(u) =  a^{2} \phi_{2}(u)+\mathcal{O}\left(a^{4}\right).
\end{eqnarray}
given the asymptotic AdS boundary conditions and horizon conditions $u=u_{h}$, one obtains
\begin{eqnarray}
\mathcal{F}_{2}(u) & = & \frac{1}{24 u_{h}^{2}}\left[8 u^{2}\left(u_{h}^{2}-u^{2}\right)-10 u^{4} \log 2+\left(3 u_{h}^{4}+7 u^{4}\right) \log \left(1+\frac{u^{2}}{u_{h}^{2}}\right)\right], \\
B_{2}(u) & = & -\frac{u_{h}^{2}}{24}\left[\frac{10 u^{2}}{u_{h}^{2}+u^{2}}+\log \left(1+\frac{u^{2}}{u_{h}^{2}}\right)\right], \\
\phi_{2}(u) & = & -\frac{u_{h}^{2}}{4} \log \left(1+\frac{u^{2}}{u_{h}^{2}}\right).
\end{eqnarray}

\begin{figure}
\centering
    \resizebox{0.6\textwidth}{!}{
    \includegraphics{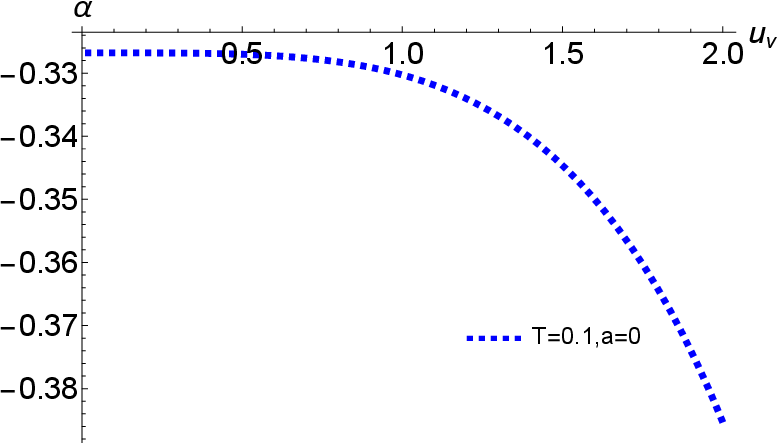}}
    \caption{\label{uvalpha} The dependence of $\alpha$ on $u_{v}$ at $T =0.1\mathrm{GeV}$ with $a=0$.  }
\end{figure}

In the context of AdS/CFT, once the metric is obtained, we can study the potential energy of three quarks. The total action for this potential energy comprises three components: the Nambu-Goto action of three quarks, the vertex action and the boundary action($S_{A}$), namely
\begin{equation}
S=\sum_{i=1}^{3} S_{N G}^{(i)}+S_{\mathrm{vert}}+3 S_{A}.
\end{equation}
Here the baryon vertex action is
\begin{equation}
S_{\mathrm{vert}}=\tau_{v} \int \mathrm{d} t \sqrt{\frac{\mathcal{F}\mathcal{B} e^{\frac{5}{2} \phi}}{u^2}}.\label{vert}
\end{equation}
Here $\tau_{v}$ is a dimensionless parameter. The  boundary  action  is  defined  as
\begin{equation}
S_{A}=\mp \frac{1}{3} \int \mathrm{d} t \mathcal{F}_{2}(0).
\end{equation}
Thus, the total action can be written as
\begin{equation}
S=\frac{3 g}{T} \int \mathrm{d} x \frac{\sqrt{\mathcal{B}}}{u^{2}} \sqrt{\mathcal{F} \mathcal{H}+\left(\partial_{x} u\right)^{2}}+\frac{3 \mathrm{~kg}}{T} \sqrt{\frac{\mathcal{F}\mathcal{B} e^{\frac{5}{2} \phi}}{u_{v}^2}}-\frac{\mathcal{F}_{2}(0)}{T}.\label{S}
\end{equation}
Following Refs.\cite{Andreev:2015riv,Jiang:2022zbt,Jiang:2023lmj}, we adopt the model-dependent parameters as follows: g=0.176 and k=-0.102.  Therefore, the Lagrange of the first term in Eq.\ref{S} is
\begin{equation}
\mathcal{L}=\frac{\sqrt{\mathcal{B}}}{u^{2}} \sqrt{\mathcal{F} \mathcal{H}+\left(\partial_{x} u\right)^{2}}.
\end{equation}
The conservation current equation is
\begin{eqnarray}
\mathcal{L}-u^{\prime} \frac{\partial \mathcal{L}}{\partial u^{\prime}}  =  \frac{\frac{\sqrt{\mathcal{B}}}{u^{2}} \mathcal{F} \mathcal{H}}{\sqrt{\mathcal{F} \mathcal{H}+\left(\partial_{x} u\right)^{2}}}  =  \text { C}.
\end{eqnarray}
So, for coordinates $u_{0}$ and vertex $u_{v}$, we have
\begin{eqnarray}
 \frac{\frac{\sqrt{\mathcal{B}(u)}}{u^{2}} \mathcal{F}(u) \mathcal{H}(u)}{\sqrt{\mathcal{F}(u)\mathcal{H}(u)+\left(\partial_{x} u\right)^{2}}} & = & \frac{\sqrt{\mathcal{B}(u_0)}}{u_0^{2}} \mathcal{F}(u_0) \mathcal{H}(u_0)\label{x0}, \\
\frac{\frac{\sqrt{\mathcal{B}(u_v)}}{u_v^{2}} \mathcal{F}(u_v)\mathcal{H}(u_v)}{\sqrt{\mathcal{F}(u_v)\mathcal{H}(u_v)+\tan ^{2} \alpha}} & = & \frac{\sqrt{\mathcal{B}(u_0)}}{u_0^{2}} \mathcal{F}(u_0) \mathcal{H}(u_0).\label{tans}
\end{eqnarray}
Note that $\partial_{x} u= \tan\alpha$ at $u=u_{v}$. Sort out the square shift terms on both sides of Eq.\ref{x0} and obtain
\begin{eqnarray}
\partial_{u} x & = & \sqrt{\frac{\frac{\mathcal{B}(u_{0})}{u_{0}^{4}} \mathcal{F}^2(u_0)\mathcal{H}^2(u_0)}{\frac{\mathcal{B}(u)}{u^{4}} \mathcal{F}^2(u)\mathcal{H}^2(u), \mathcal{F}(u_0)\mathcal{H}(u_0)-\frac{\mathcal{B}(u_{0})}{u_{0}^{4}} \mathcal{F}^2(u_0)\mathcal{H}^2(u_0) \mathcal{F}(u)\mathcal{H}(u)}}.
\end{eqnarray}
Then multiply the denominator on both sides of the Eq.\ref{tans} to obtain
\begin{eqnarray}
\frac{\sqrt{\mathcal{B}(u_v)}}{u_{v}^{2}} \mathcal{F}(u_v)\mathcal{H}(u_v)-\frac{\sqrt{\mathcal{B}(u_0)}}{u_{0}^{2}} \sqrt{\mathcal{F}(u_0)\mathcal{H}(u_0)}, \sqrt{\mathcal{F}(u_v)\mathcal{H}(u_v)+\tan ^{2} \alpha}=0.
\end{eqnarray}
From Fig.\ref{A}, the  force  balance  equation  at  $u = u_{v}$ is
\begin{eqnarray}
\boldsymbol{e}_{1}+\boldsymbol{e}_{2}+\boldsymbol{e}_{3}+\boldsymbol{f}_{v} & = & 0.
\end{eqnarray}
Here $\boldsymbol{e}_{1}$, $\boldsymbol{e}_{2}$ and $\boldsymbol{e}_{3}$  are string tensions. $\boldsymbol{f}_{v}$  is  gravitational
force at the  vertex, which is given by $f_{v}=-\delta E_{\mathrm{vert}} / \delta u$ with $E_{\mathrm{vert}} =TS_{vert}$. Where $S_{vert}$ is the action of the vertex, which can be given by Eq.\ref{vert}.  Therefore each force and its component is
\begin{eqnarray}
f_{v} & = & \left(0,0,-3 g \mathrm{k} \partial_{u_{v}} \sqrt{ \frac{ \mathcal{F}(u_v) \mathcal{B}(u_v) e^{ \frac{5}{2} \phi }}{u^2} }\right),\notag\\
\boldsymbol{e}_{\mathbf{1}} & = & -g \frac{\sqrt{\mathcal{B}(u_v)}}{u_{v}^{2}}\left(\cos \beta \frac{\mathcal{F}(u_v) \mathcal{H}(u_v)}{\sqrt{\mathcal{F}(u_v) \mathcal{H}(u_v)+\tan ^{2} \alpha}},
\sin \beta \frac{\mathcal{F}(u_v) \mathcal{H}(u_v)}{\sqrt{\mathcal{F}(u_v) \mathcal{H}(u_v)+\tan ^{2} \alpha}}\right.,\notag\\
&-& \left.\frac{1}{\sqrt{1+\mathcal{F}(u_v) \mathcal{H}(u_v) \cot ^{2} \alpha}}\right)\notag\\
\boldsymbol{e}_{2} & = & -g \frac{\sqrt{\mathcal{B}(u_v)}}{u_{v}^{2}}\left(-\cos \beta \frac{\mathcal{F}(u_v) \mathcal{H}(u_v)}{\sqrt{\mathcal{F}(u_v) \mathcal{H}(u_v)+\tan ^{2} \alpha}},\sin \beta \frac{ \mathcal{F}(u_v) \mathcal{H}(u_v) }{\sqrt{ \mathcal{F}(u_v) \mathcal{H}(u_v)+\tan^{2} \alpha}}\right.,\notag\\
 &-& \left.\frac{1}{\sqrt{1+\mathcal{F}(u_v) \mathcal{H}(u_v) \cot ^{2} \alpha}}\right),\notag \\
\boldsymbol{e}_{3} & = & -g \frac{\sqrt{\mathcal{B}(u_v)}}{u_{v}^{2}}\left(0,-\frac{\mathcal{F}(u_v) \mathcal{H}(u_v)}{\sqrt{\mathcal{F}(u_v) \mathcal{H}(u_v)+\tan ^{2} \alpha}},-\frac{1}{\sqrt{1+\mathcal{F}(u_v) \mathcal{H}(u_v) \cot ^{2} \alpha}}\right) \text {. } \notag\\
\end{eqnarray}
Due to the fact that the forces acting in the $x$ and $y$ directions are equal in magnitude and opposite in direction, the combined external force acting is directly equal to zero. Then, according to the force  balance equation, we have
\begin{eqnarray}
\frac{\sqrt{\mathcal{B}(u_v)}}{u_{v}^{2}} \frac{1}{\sqrt{1+\mathcal{F}(u_v) \mathcal{H}(u_v) \cot ^{2} \alpha}}-k \partial_{u_{v}} \sqrt{ \frac{ \mathcal{F}(u_v) \mathcal{B}(u_v) e^{ \frac{5}{2} \phi }}{u_v^2} } & = & 0.\label{uv}
\end{eqnarray}
By solving Eq.\ref{uv}, $u_{v}$ can be directly obtained. In the A configuration, the length of the strings connecting the three quarks can be viewed as an integral from 0 to $u_{0}$ and then to $u_{v}$. Therefore, we have the following expression\cite{Jiang:2022zbt,Andreev:2015riv}
\begin{eqnarray}
L & = & \sqrt{3}\left(\int_{0}^{u_{0}} \partial_{u} x \mathrm{~d} u+\int_{u_{v}}^{u_{0}} \partial_{u} x \mathrm{~d} u\right)\notag \\ & = & \sqrt{3} \int_{0}^{u_{0}} \sqrt{\frac{\frac{\mathcal{B}(u_0)}{u_{0}^{4}} \mathcal{F}^2(u_0) \mathcal{H}^2(u_0)}{\frac{\mathcal{B}(u)}{u^{4}} \mathcal{F}^2(u) \mathcal{H}^2(u) \mathcal{F}(u_0) \mathcal{H}(u_0))-\frac{\mathcal{B}(u_0)}{u_{0}^{4}} \mathcal{F}^2(u_0) \mathcal{H}^2(u_0) \mathcal{F}(u) \mathcal{H}(u)}} d u \notag\\
&+& \sqrt{3} \int_{u_{v}}^{v_{0}} \sqrt{\frac{\frac{\mathcal{B}(u)}{r_{0}^{4}} \mathcal{F}^2(u_0) \mathcal{H}^2(u_0)}{\frac{\mathcal{B}(u)}{u^{4}} \mathcal{F}^2(u) \mathcal{H}^2(u) \mathcal{F}(u_0) \mathcal{H}(u_0)-\frac{\mathcal{B}(u_0)}{u_{0}^{4}} \mathcal{F}^2(u_0) \mathcal{H}^2(u_0) \mathcal{F}(u) \mathcal{H}(u)}} \mathrm{d} u.\notag \\
\end{eqnarray}
Thus the potential energy is
\begin{eqnarray}
E & = & 3 g\left(\int_{0}^{u_{0}} \frac{\sqrt{\mathcal{B}(u)}}{u^{2}} \sqrt{1+\mathcal{F}(u) \mathcal{H}(u)\left(\partial_{u} x\right)^{2}}-\frac{1}{u^{2}} \mathrm{~d} u\right.\notag\\
&+&\left.\int_{u_{v}}^{u_{0}} \frac{\sqrt{\mathcal{B}(u)}}{u^{2}} \sqrt{1+\mathcal{F}(u) \mathcal{H}(u)\left(\partial_{u} x\right)^{2}} \mathrm{~d} u\right) -\frac{3 g}{u_{0}}+3 g \mathrm{k} \sqrt{ \frac{ \mathcal{F}(u_v) \mathcal{B}(u_v) e^{ \frac{5}{2} \phi }}{u_v^2} }+3 c \notag\\
&-&\mathcal{F}_{2}(0).
\end{eqnarray}
Here we take c=0.623GeV.
Based on the theoretical derivation and discussion above, we turn to analyzing its calculation results. Firstly, let's focus on vertex coordinate.  The relationship between angle $\alpha$ and the vertex coordinate $u_{v}$ can be described by the balance equation, as shown in Fig.\ref{uvalpha}. It is found that this relationship represents a monotonic function. Specifically, when $u_{v}$ has a very small magnitude, the change in angle $\alpha$ is relatively gradual. This implies that $\alpha$ exhibits low sensitivity to variations in $u_{v}$ under such conditions. However, when $u_{v}$ takes on larger values, the angle $\alpha$ undergoes rapid and significant changes in response to the fluctuations in $u_{v}$.
\begin{figure}
\centering
    \resizebox{0.6\textwidth}{!}{
    \includegraphics{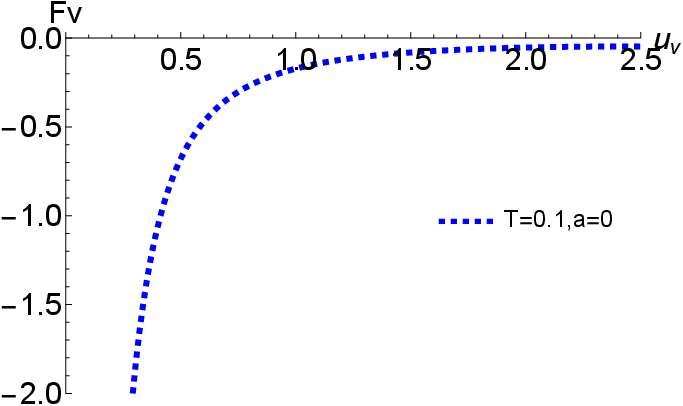}}
    \caption{\label{FuvA} The dependence of $F_{v}$ on $u_{v}$ at $T =0.1\mathrm{GeV}$ }
\end{figure}

\begin{figure}
\centering
    \resizebox{0.6\textwidth}{!}{
    \includegraphics{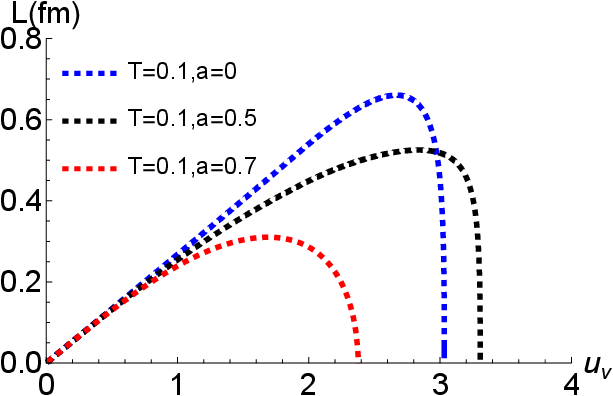}}
    \caption{\label{Luv} The  dependence  of  separation  distance  of  triply  heavy  baryon   on $u_{v}$ at $T =0.1\mathrm{GeV}$ with $a$. The blue dashed line is $a=0$, the black line is $a=0.5$ and the red line is $a=0.7$. }
\end{figure}

In order to further investigate the forces $F_{v}$ affecting the vertex, we plot Fig.\ref{FuvA}. It reveals a consistent trend where the magnitude of the force acting on the vertex decreases gradually as the $u_{v}$ increases.
In this model, the three quarks are interconnected by strings, and we are aware that the separation distance of heavy quarkonium varies in response to changes in the anisotropic background. This leads us to investigate how the separation distance among the three quarks is affected. In Fig.\ref{Luv}, we present the variations in separation distance and the parameter $u_{v}$ for the three quarks. It is found that the separation distance between the three quarks steadily increases as $u_{v}$ increases, reaching its maximum value before gradually decreasing. This observed behavior closely resembles the response of heavy quarkonium. Furthermore, it's worth noting that as the anisotropy parameter $a$ progressively increases, the maximum screening distance consistently decreases. This indicates that $a$ has a diminishing effect on the screening distance.

\begin{figure}
\centering
    \resizebox{0.6\textwidth}{!}{
    \includegraphics{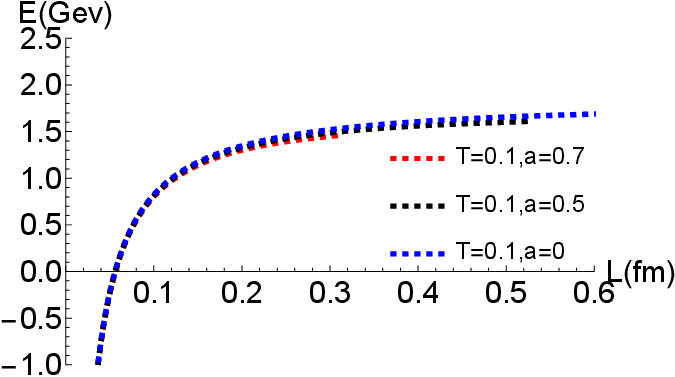}}
    \caption{\label{LEA} The dependence of potential  of  triply  heavy  baryon  on  separation  distance at $T =0.1\mathrm{GeV}$ with a. The blue dashed line is $a=0$, the black line is $a=0.5$ and the red line is $a=0.7$. }
\end{figure}

To investigate the dependency of the potential energy of the three quarks on the separation  distance $L$, we plot Fig.\ref{LEA}. The results show that the potential energy of the three quarks exhibits a continuous increase with the separation distance, ultimately reaching a point where the three quark systems undergo dissolution. Besides, the results also indicate that with an increase in anisotropy, the potential energy of the three quarks gradually diminishes. This signifies that the anisotropic parameters exert a suppressing effect on the potential energy. In particular, the impact of the anisotropy parameter $a$ is most pronounced in the linear region of the potential energy.
\section{Three-quark potential of B configuration}\label{sec:03_Three-quark potential of B model}
\begin{figure}[H]
\centering
    \resizebox{0.6\textwidth}{!}{
    \includegraphics{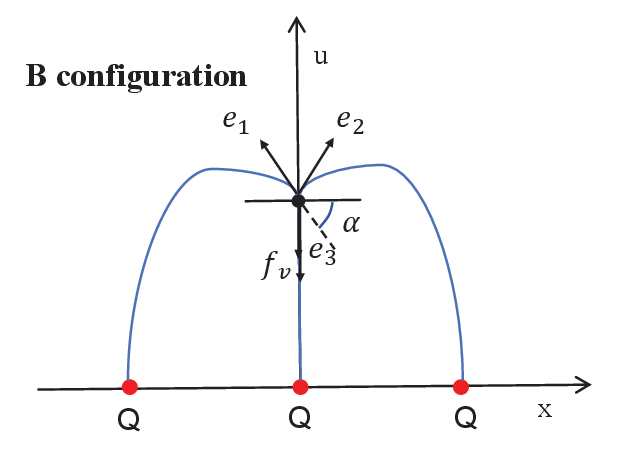}}
    \caption{\label{B}The schematic diagram of B configuration. Red dots are employed to represent heavy quarks, black dot is used to denote verter, and solid blue lines are utilized to illustrate strings. The system is anisotropic in the transverse direction.}
\end{figure}

In the preceding section, we focus on the separation distance and potential energy of three quarks within the A configuration. Now, let's turn our attention to B configuration, where one quark is positioned at the coordinate origin, and the other two quarks are symmetrically distributed on both sides of the x-axis, as shown in Fig.\ref{B}. The system is anisotropic in the transverse direction. Thus, the force balance equation for the B configuration is
\begin{eqnarray}
-2 \frac{ \sqrt{\mathcal{B}(u_v)} }{u_{v}^{2}} \frac{1}{\sqrt{1+\mathcal{F}(u_v) \mathcal{H}(u_v) \cot ^{2} \alpha}}+\frac{  \sqrt{\mathcal{B}(u_v)}  }{u_{v}^{2}}+3 k \partial_{u_{v}} \sqrt{ \frac{ \mathcal{F}(u_v) \mathcal{B}(u_v) e^{ \frac{5}{2} \phi }}{u^2} } & = & 0.\label{Buv}
\end{eqnarray}
Then, the vertex coordinate $u_{v}$ can be obtained by applying Eq.\ref{Buv}. Additionally, the value of $u_{0}$ can be determined through Eq.\ref{tans} as well. By utilizing these obtained values, we can calculate the separation distance of B configuration
\begin{eqnarray}
L & =&  \int_{0}^{u_{0}} \partial_{u} x \mathrm{~d} u+\int_{u_{v}}^{u_{0}} \partial_{u} x \mathrm{~d} u \notag\\ & = & \int_{0}^{u_{0}} \sqrt{\frac{\frac{\mathcal{B}(u_0)}{u_{0}^{4}} \mathcal{F}^2(u_0) \mathcal{H}^2(u_0) }{\frac{\mathrm{e}^{2 s r^{2}}}{u^{4}} \mathcal{F}^2(u) \mathcal{H}^2(u) \mathcal{F}(u_0) \mathcal{H}(u_0)-\frac{\mathcal{B}(u_0)}{u_{0}^{4}} \mathcal{F}^2(u_0) \mathcal{H}^2(u_0) \mathcal{F}^2(u) \mathcal{H}^2(u)}} \mathrm{d} u\notag\\
&+&\int_{u_{v}}^{u_{0}} \sqrt{\frac{\frac{\mathcal{B}(u_0)}{u_{0}^{4}} \mathcal{F}^2(u_0) \mathcal{H}^2(u_0)}{\frac{\mathcal{B}(u)}{u^{4}} \mathcal{F}^2(u) \mathcal{H}^2(u)\mathcal{F}(u_0) \mathcal{H}(u_0)-\frac{\mathcal{B}(u_0)}{u_{0}^{4}} \mathcal{F}^2(u_0) \mathcal{H}^2(u_0) \mathcal{F}(u) \mathcal{H}(u)}} \mathrm{d} u. \notag\\
\end{eqnarray}
Thus, the potential energy is given by
\begin{eqnarray}
E &=&  2 g\left(\int_{0}^{u_{0}} \frac{\mathcal{B}(u)}{u^{2}} \sqrt{1+\mathcal{F}(u) \mathcal{H}(u)\left(\partial_{u} x\right)^{2}} \mathrm{~d} u\right.\notag\\
&+&\left.\int_{u_{v}}^{u_{0}} \frac{\mathcal{B}(u)}{u^{2}} \sqrt{1+\mathcal{F}(u) \mathcal{H}(u)\left(\partial_{u} x\right)^{2}} \mathrm{~d} u\right)
-\frac{2 g}{u_{0}}+g \int_{0}^{u_{v}} \left(\frac{\mathcal{B}(u)}{u^{2}}-\frac{1}{u^{2}}\right) \mathrm{~d} u-\frac{g}{u_{v}}\notag\\
 &+&3 g k \sqrt{\frac{\mathcal{F}(u) \mathcal{B}(u) e^{\frac{5}{2} \phi}}{u^2}}+3 c-\mathcal{F}_{2}(0).
\end{eqnarray}

\begin{figure}
\centering
    \resizebox{0.6\textwidth}{!}{
    \includegraphics{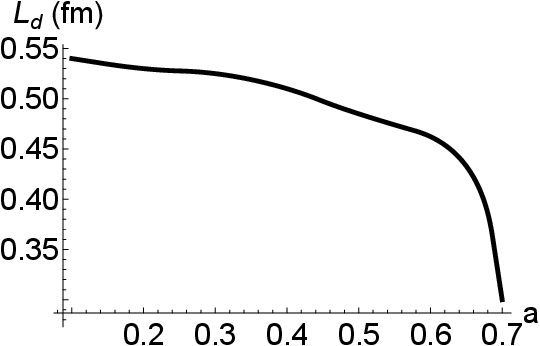}}
    \caption{\label{LdB} The dependence of screening distance on $a$ at $T =0.1\mathrm{GeV}$.}
\end{figure}

The physical significance of the screening distance is its capability to provide insights into the stability and dissolution characteristics of three-quark systems within strongly interacting media. In particular, when the screening distance is substantial, it signifies that the quarkonium remains relatively stable within the surrounding medium and can maintain its original state. Thus, to investigate the stability of the three-quark system, we plot Fig.\ref{LdB}. The result of this analysis reveal that, as the parameter $a$ increases, the maximum screening distance gradually diminishes, indicating that the three-quark system becomes more susceptible to dissolution under these conditions.

 \begin{figure}
 \centering
    \resizebox{0.6\textwidth}{!}{
    \includegraphics{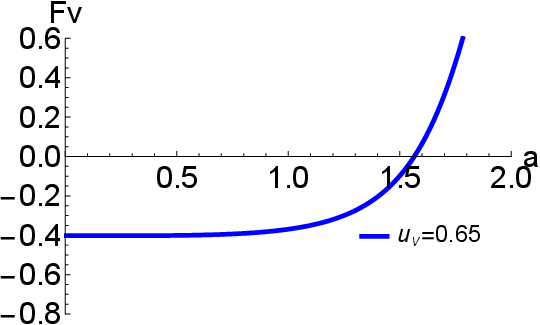}}
    \caption{\label{FvaB} The dependence of $F_{v}$ on $a$ at $T =0.1\mathrm{GeV}$.}
\end{figure}

To further investigate the force acting on vertex, we plot Fig.\ref{FvaB}. The results indicate that as the anisotropy parameter increases, the force acting on the vertex can change from negative to positive. A detailed explanation of this behavior can be provided as follows: As shown in  Fig.\ref{B}, the force acting on the vertex is clearly directed downward. Thus, a positive force value typically corresponds to an upward force direction. However, considering the force acting on the vertex to be positive in this particular context would imply a direction contrary to that depicted in Fig.\ref{B}. This discrepancy clearly lacks physical significance and contradicts the established force direction.
Therefore, we can conclude that when the anisotropic parameter $a$ exceeds a certain critical value $a_{0}$ ($a > a_{0}$, where $a_{0}$ represents the root of equation $F_{v}(a)=0$), the three quark system will become unstable.

\begin{figure}
\centering
    \resizebox{0.6\textwidth}{!}{
    \includegraphics{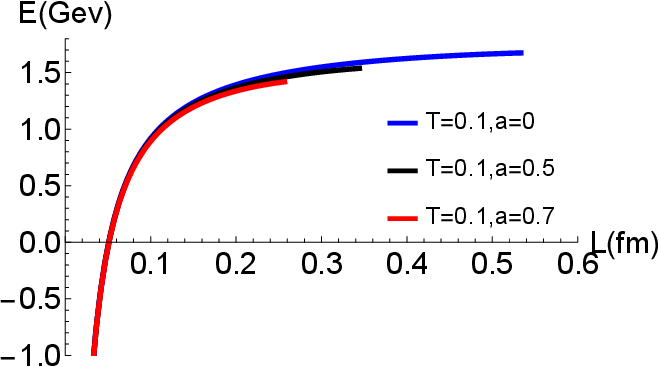}}
    \caption{\label{LEB} The dependence of $\alpha$ on $u_{v}$ at $T =0.1\mathrm{GeV}$ with a. The blue line is $a=0$, the black line is $a=0.5$ and the red line is $a=0.7$. }
\end{figure}

\begin{figure}
\centering
    \resizebox{0.6\textwidth}{!}{
    \includegraphics{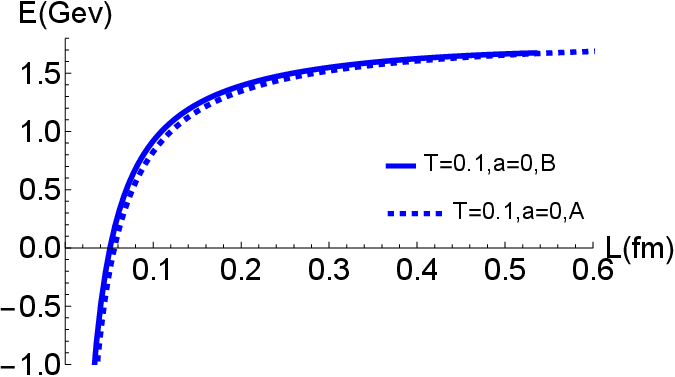}}
    \caption{\label{EAVSEB} Comparing the potential energy in A configuration and B configuration at $T=0.1GeV$ and $a=0$. The blue dashed line is A configuration and the blue line is B configuration. }
\end{figure}

The Fig.\ref{LEB} illustrates the correlation between potential energy (E) and separation distance (L). In general, the potential energy consistently rises as the separation distance increases. Additionally, as the parameter $a$ increases, there is a gradual decrease in potential energy. These two observations closely parallel the findings in A configuration. However, the energy of the three quarks in A configuration is slightly lower than that in B configuration, shown in Fig.\ref{EAVSEB}. This difference in potential energy  implies that the three quarks in A configuration have relatively greater stability compared to those in B configuration, which is consistent with the results of Ref.\cite{Jiang:2022zbt}

\section{Three-quark potential with the same entropy density}\label{sec:04}

As it was expressed, to compare the results with the isotropic case one can fix temperature and change entropy density. In Ref.\cite{BitaghsirFadafan:2017tci}, it was found that the behavior of a jet at a specific temperature is distinct from its behavior at the same entropy density. In the preceding section, we studied the potential energy of three quarks under identical temperature conditions. Now, our focus shifts to investigating the potential energy of three quarks subjected to an equivalent energy density. Here, the entropy density can be determined using Eq.\ref{s}, and the method for calculating the $L$ and $E$ of the three-quark system under same entropy density conditions is analogous to the previous calculations.

\begin{figure}[H]
    \resizebox{1\textwidth}{!}{
    \includegraphics{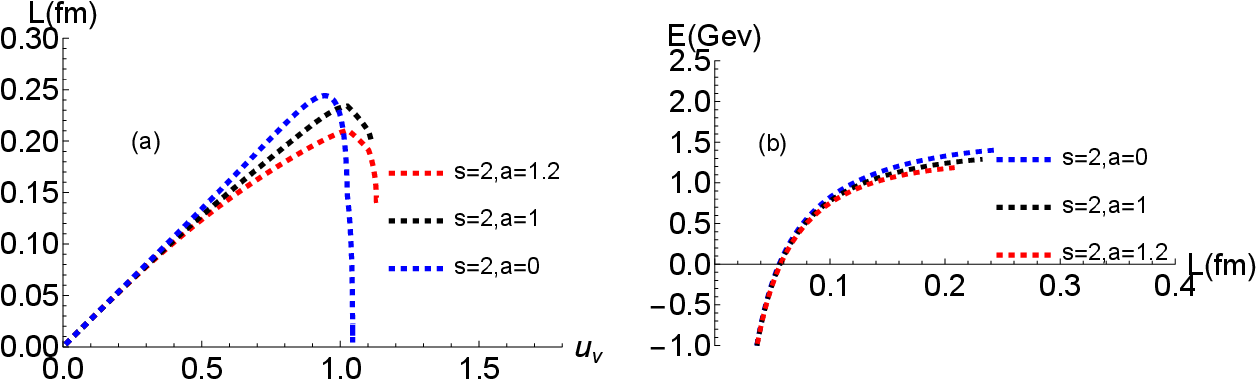}}
    \caption{\label{SLE} (a)The dependence of $L$ on $u_{v}$ at $s =2$ with $a$ for A configuration. The blue dashed line is $a=0$, the black line is $a=1$ and the red line is $a=1.2$. (b)The dependence of $E$ on $L$ at $s =2$ with $a$. The blue dashed line is $a=0$, the black line is $a=1$ and the red line is $a=1.2$. }
\end{figure}

\begin{figure}[H]
    \resizebox{1\textwidth}{!}{
    \includegraphics{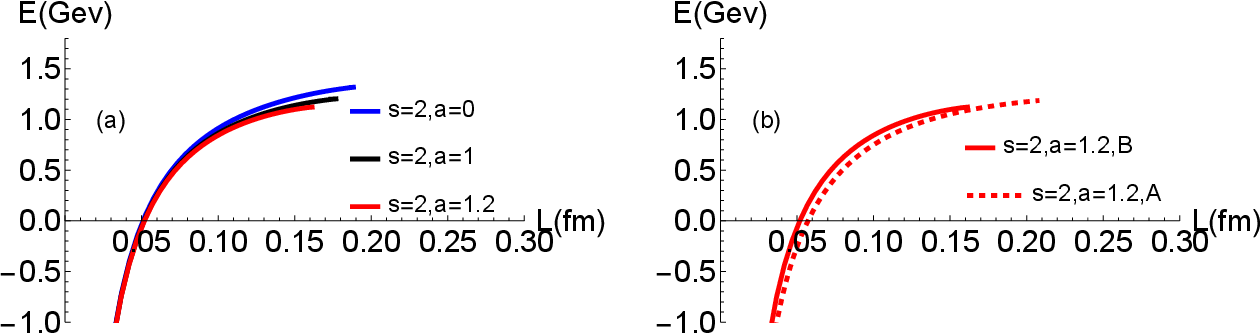}}
    \caption{\label{SLEE} (a)The dependence of $E$ on $L$ at $s =2$ with $a$ for B configuration. The blue line is $a=0$, the black line is $a=1$ and the red line is $a=1.2$. (b)Comparing the potential energy in A configuration and B configuration at $s =2$ and $a=1.2$. The red dashed line is A configuration and the red line is B configuration. }
\end{figure}

Based on the results presented in Fig.\ref{SLE}, it is found that in A configuration, where the entropy density remains constant, the maximum screening distance steadily decreases as the anisotropy parameter $a$ increases. This means that the presence of anisotropic parameters is beneficial for the dissolution of three-quark systems when the entropy density or degree of system disorder is equivalent. The potential energy at fixed entropy density also exhibits a continuous decline with increasing $a$, which is very similar to the behavior of fixed temperature, as shown in Fig.\ref{SLEE}(a). However, the distinction emerges in terms of the potential energy between A configuration and B configuration under fixed entropy density, with the former exhibiting significantly lower values than the latter(Fig.\ref{SLEE}(b)). This also means that the three quark system in A configuration is more stable than in B configuration.

\section{Summary and Conclusions}\label{sec:05}
In this work, we mainly study triply heavy baryon in anisotropic backgrounds. The total action of the three quark system mainly includes the Nambu-Goto action, vertex action, and boundary term of the three quarks. The vertex positions can be obtained by solving the force
balance equation. After determining the vertex position, we further calculate the separation distance and potential energy of the three quarks. The results show that as the anisotropic background $a$ increases, the maximum screening distance gradually decreases. Next, we also study the three quark potential energy in A configuration and B configuration. The results indicate that in both A configuration and B configuration, the potential energy gradually decreases with the increase of anisotropic parameter $a$. This shows that the presence of an anisotropic background promotes the dissolution of the three quark system. The difference is that the potential energy in A configuration is smaller than that in B configuration, which means that the three quark system in A configuration is more stable. That would be interesting to add chemical potential and study holographic A and B configuration of heavy baryons in the presence of anisotropy \cite{muaniso}.

However, triple heavy baryons have not been directly observed in experiments. Therefore, our work can be seen as a prediction of experimental. Over the past decade, many exotic hadron states have been found through experiments. These exotic hadron states offer a new perspective for understanding the dynamics of strong interactions and multi-quark systems. Specifically, $X(6900)$ has been observed on the LHC and is assumed to be composed of four fully heavy quarks\cite{Deng:2020iqw}. But so far, its properties are not entirely clear. Therefore, it is necessary for us to study the potential energy of fully heavy tetraquark in anisotropic backgrounds, in order to broaden our understanding of exotic hadron states. We will further investigate it in the future work.
\section*{Acknowledgments}
This work is partly supported by the Natural Science Foundation of Hunan Province, China(Grant No.2021JJ40020) and the Research Foundation of Education Bureau of Hunan Province, China under Grant Nos. 21B0402 and 22B0788. KBF would like to thank UCAS for hospitality.

\end{document}